\documentclass[sn-mathphys,Numbered]{sn-jnl}

\usepackage{graphicx}%
\usepackage{multirow}%
\usepackage{amsmath,amssymb,amsfonts}%
\usepackage{amsthm}%
\usepackage{mathrsfs}%
\usepackage[title]{appendix}%
\usepackage{xcolor}%
\usepackage{textcomp}%
\usepackage{manyfoot}%
\usepackage{booktabs}%
\usepackage{algorithm}%
\usepackage{algorithmicx}%
\usepackage{algpseudocode}%
\usepackage{listings}%

\theoremstyle{thmstyleone}%
\theoremstyle{thmstyletwo}%

\theoremstyle{thmstylethree}%

\begin{document}

\title[Effect of annealing in eutectic high-entropy alloy superconductor NbScTiZr]{Effect of annealing in eutectic high-entropy alloy superconductor NbScTiZr}


\author[1]{\fnm{Takeru} \sur{Seki}}

\author[2]{\fnm{Hiroto} \sur{Arima}}

\author[3]{\fnm{Yuta} \sur{Kawasaki}}

\author[3]{\fnm{Terukazu} \sur{Nishizaki}}

\author[2]{\fnm{Yoshikazu} \sur{Mizuguchi}}

\author*[1]{\fnm{Jiro} \sur{Kitagawa}}\email{j-kitagawa@fit.ac.jp}

\affil[1]{\orgdiv{Department of Electrical Engineering, Faculty of Engineering}, \orgname{Fukuoka Institute of Technology}, \orgaddress{\street{3-30-1 Wajiro-higashi, Higashi-ku}, \city{Fukuoka}, \postcode{811-0295}, \country{Japan}}}

\affil[2]{\orgdiv{Department of Physics}, \orgname{Tokyo Metropolitan University}, \orgaddress{\street{1-1, Minami-osawa}, \city{Hachioji}, \postcode{192-0397}, \country{Japan}}}

\affil[3]{\orgdiv{Department of Electrical Engineering, Faculty of Science and Engineering}, \orgname{Kyushu Sangyo University}, \orgaddress{\street{2-3-1 Matsukadai, Higashi-ku}, \city{Fukuoka}, \postcode{813-8503}, \country{Japan}}}


\abstract{We investigated the impact of annealing on the structural characteristics and superconducting critical temperature ($T_\mathrm{c}$) of the eutectic high-entropy alloy (HEA) superconductor NbScTiZr. The HEA manifests an eutectic microstructure composed of body-centered cubic (bcc) and hexagonal close-packed phases. Both the lattice parameters of the bcc phase and grain size of the eutectic structure exhibited pronounced sensitivity to variations in annealing temperature. The observed dependence of the lattice parameter on annealing temperature supports the possibility that lattice strain occurs at lower annealing temperatures. The as-cast sample demonstrated superconductivity at $T_\mathrm{c}$ of 7.9 K, which increased to 9 K after annealing at 800 $^{\circ}$C. However, when subjected to annealing at 1000 $^{\circ}$C, $T_\mathrm{c}$ diminishes to 8.7 K. The annealing-temperature dependence of $T_\mathrm{c}$ cannot be comprehensively elucidated based solely on the electronic density of states at the Fermi level. It is plausible that the lattice strain may influence the annealing temperature dependence of $T_\mathrm{c}$. Our results for the critical current density $J_{c}$ reveal that the self-field $J_{c}$ of the as-cast NbScTiZr at 2 K exceeds 10$^{6}$ A/cm$^{2}$.}




\maketitle

\section{Introduction}\label{sec1}
High-entropy alloys (HEA) constitute a new research area in materials science, and this innovative domain has been extensively applied to a diverse array of alloys and crystalline systems\cite{Biswas:book}.
Unlike their conventional counterparts, which comprise one or two principal elements, HEAs generally contain at least four principal elements\cite{Lu:SM2020,Yan:MMTA2021}. 
HEAs have attracted considerable attention as promising candidates for structural materials owing to their excellent mechanical properties\cite{Li:PMS2021}. 
In addition, HEAs exhibit numerous versatile functionalities, including energy storage, magnetic refrigeration, soft ferromagnetism, catalysis, thermoelectricity, and biocompatibility\cite{Marques:EES2021,Yang:NM2022,Law:JMR2023,Nakamura:AIPAd2023,Kitagawa:APLMater2022,Kitagawa:JMMM2022,Wu:JACS2020,Jiang:Science2021,Castro:Metals2021}.
Another feature in HEAs is functionality augmentation by manipulating microstructure, such as eutectic structures.
Notably, instances arise in which the microstructural features exert a discernible influence on the manifested physical properties.\cite{Lu:SM2020,Bhardwaj:TI2021,Rao:AFM2021,Baba:Materials2021}. 
One extensively studied area is structural HEA materials, with eutectic HEAs capturing particular interest for their potential to surmount the traditional strength-ductility trade-off encountered in structural materials.\cite{Lu:SM2020,Bhardwaj:TI2021}. 
Body-centered cubic (bcc) HEAs exhibit limited ductility, whereas face-centered-cubic (fcc) HEAs exhibit high ductility and low strength.
Accordingly, the pursuit of concurrent high strength and ductility is realized by strategically including bcc and fcc phases within eutectic HEAs.
Another example is a magnetic HEA\cite{Rao:AFM2021} denoted as Fe$_{15}$Co$_{15}$Ni$_{20}$Mn$_{20}$Cu$_{30}$.
This particular alloy exhibited spinodal decomposition after thermal treatment, and controlling the spinodal structure by changing the annealing temperature increased the Curie temperature.

HEA superconductivity has been extensively investigated \cite{Kozelj:PRL2014,Sun:PRM2019,Kitagawa:Metals2020} since the discovery of bcc Ta$_{34}$Nb$_{33}$Hf$_{8}$Zr$_{14}$Ti$_{11}$. 
The HEA superconductors are documented in various structural types, including bcc\cite{Rohr:PRM2018,Marik:JALCOM2018,Ishizu:RINP2019,Harayama:JSNM2021,Sarkar:IM2022,Motla:PRB2022,Kitagawa:RHP2022,Kitagawa:JALCOM2022,Hattori:JAMS2023,Li:JPCC2023,Zeng:SCPMA2023}, hexagonal close-packed (hcp)\cite{Lee:PhysicaC2019,Marik:PRM2019,Browne:JSSC2023}, CsCl-type\cite{Stolze:ChemMater2018}, NaCl-type\cite{Mizuguchi:JPSJ2019,Yamashita:DalTran2020}, $\alpha$ (or $\beta$)-Mn-type\cite{Stolze:JMCC2018,Xiao:SM2023}, CuAl$_{2}$-type\cite{Kasen:SST2021}, W$_{5}$Si$_{3}$-type\cite{Liu:PRM2023}, BiS$_{2}$-based, and YBCO-based\cite{Sogabe:SSC2019,Shukunami:PhysicaC2020} structures.
Several interesting features of the HEAs have been identified. 
The (TaNb)$_{0.67}$(HfZrTi)$_{0.33}$ HEA with a bcc structure exhibits a robustness of superconductivity under high pressures\cite{Guo:PNAC2017}.
Remarkably, the superconducting critical temperature ($T_\mathrm{c}$) remained nearly invariant at 8 K, even when subjected to an extreme pressure of 180 GPa.
The "cocktail effect", whereby the physical properties of an HEA surpass those of its constituent elements, is a crucial aspect of HEA materials. 
This effect is evident in BiS$_{2}$-based superconductors, where the diamagnetic signal is enhanced within the high-entropy state\cite{Sogabe:SSC2019}.
A recent report revealed that the critical current density $J_{c}$ of a Hf-Nb-Ta-Ti-Zr film exceeds that of an NbTi alloy employed in high-field superconducting magnets\cite{Jung:NC2022}. 
Emerging research has underscored the notable inverse correlation between $T_\mathrm{c}$ and the Debye temperature across several bcc HEA superconductors\cite{Kitagawa:JALCOM2022,Hattori:JAMS2023}.
This behavior can be understood using the Bardeen–Cooper–Schrieffer (BCS) theory, which considers the uncertainty principle between the phonon energy and phonon lifetime\cite{Kitagawa:JALCOM2022,Hattori:JAMS2023}.

In several superconductors, the inclusion of eutectic phases affects the superconducting properties.
A salient exemplar of this phenomenon can be discerned in Sr$_{2}$RuO$_{4}$, where the eutectic composite exhibiting a lamellar arrangement with Ru metal shows a $T_\mathrm{c}$ of 3 K, a notable advancement from the 1.5 K exhibited by pure Sr$_{2}$RuO$_{4}$\cite{Maeno:PRL1998}.
A similar effect was observed for Ir \cite{Matthias:Science1980}, which formed a eutectic phase with a small amount of YIr$_{2}$.
In this case, a strain-induced lattice softening enhances the $T_\mathrm{c}$ from 0.1 to 2.7 K\cite{Matthias:Science1980}.
Our investigations yielded parallel findings, wherein $T_\mathrm{c}$ enhancement in eutectic Zr$_{5}$Pt$_{3}$O$_{x}$ has been documented\cite{Hamamoto:MRX2018}.
In this compound, a eutectic phase comprising Zr$_{5}$Pt$_{3}$O$_{0.5-0.6}$ and ZrPt emerged for $x$ values exceeding 1.0. 
The area of the eutectic phase increases as $x$ is increased above 1.0, and $T_\mathrm{c}$ systematically increases from 3.2 K for $x$=0.6 to 4.8 K for $x$=2.5.
In eutectic HEAs, an alteration in the superconducting properties is expected by controlling the morphology of the eutectic structure.
However, exploring eutectic HEA superconductors remains a relatively nascent endeavor.
In addition, we consider $J_\mathrm{c}$ within the eutectic HEAs.
A eutectic structure possesses many grain boundaries that serve as effective magnetic flux pinning sites.
Therefore, it is conceivable that eutectic HEAs may confer a heightened
$J_\mathrm{c}$.
The practical superconducting wire employs a multifilamentary structure, and a eutectic HEA superconductor can be conceptualized as a built-in multifilamentary configuration.
Therefore, eutectic HEA superconductors are good candidates for high-performance superconducting-wire applications\cite{Kitagawa:Metals2020}. 

Our attention converges on NbScTiZr, which is a eutectic HEA composed of (NbTiZr)-enriched bcc and (ScZr)-enriched hcp phases\cite{Rogal:MSEA2016,Krnel:Materials2022}.
The superconducting properties of as-cast NbScTiZr were investigated by Krnel et al.\cite{Krnel:Materials2022}, revealing a bulk superconductivity showing $T_\mathrm{c}$ of 7.3 K.
Superconductivity occurs in the bcc phase\cite{Krnel:Materials2022}.
Although the superconductivity of the as-cast sample was investigated, the annealing effect was not studied.
We investigated the effect of annealing on $T_\mathrm{c}$ in NbScTiZr by measuring the magnetization and electrical resistivity. 
The annealing temperature dependence of $T_\mathrm{c}$ was discussed considering the structural properties and electronic structure calculations.
Furthermore, we show the results for $J_\mathrm{c}$.
Our new findings include $T_\mathrm{c}$ enhancement by thermal annealing in the eutectic HEA and a notably high $J_\mathrm{c}$ in the as-cast sample.

\section{Materials and Methods}\label{sec2}
The as-cast sample (2.5 g) was synthesized using a homemade arc furnace operating in an Ar atmosphere employing an Nb wire (Nilaco, 99.9 \%), Sc chips (Kojundo Chemical Laboratory, 99 \%), Ti wire (Nilaco, 99.9 \%), and Zr wire (Nilaco, 99.5 \%).
The atomic ratio was Nb:Sc:Ti:Zr=1:1:1:1.
Multiple iterations of sample flipping and remelting ensured sample homogeneity, and the sample was quenched on a water-cooled Cu hearth.
Subsequent heat treatment at 400 $^{\circ}$C, 800 $^{\circ}$C or 1000 $^{\circ}$C for 4 d was executed within an electronic furnace, with the sample encapsulated within an evacuated quartz tube.

X-ray diffraction (XRD) patterns were recorded using an X-ray diffractometer (Shimadzu XRD-7000L) with Cu-K$\alpha$ radiation in the Bragg-Brentano geometry. 
Thin slabs cut from the samples were used because of the difficulty in obtaining fine powders.
Scanning electron microscopy (SEM) images were captured using field-emission SEM (FE-SEM; JEOL JSM-7100F). 
An energy-dispersive X-ray (EDX) spectrometer co-located with the FE-SEM was employed to gain insight into the elemental composition.
The microstructures were observed on the polished surfaces.
The surface of each sample was polished using silicon carbide paper (\#240, \#400, \#600, and \#1000 mesh), alumina paste (5, 1, and 0.1 $\mu$m), and colloidal silica (0.04 $\mu$m).

The temperature-dependent behavior of ac magnetic susceptibility $\chi_{ac}$ ($T$) under a 5 Oe ac field at 800 Hz was assessed using a conventional mutual inductance technique facilitated by a GM refrigerator (UW404, Ulvac Cryogenics) operating within a temperature interval of 3 to 20 K.
Samples weighing $\sim$ 20 mg were used for this measurement.
The temperature dependence of dc magnetization $M$($T$) and isothermal magnetization curve spanning from -7 to 7 T were measured using a SQUID magnetometer (MPMS3, Quantum Design).
To measure $M$, we prepared samples with dimensions of $\sim$2$\times$1$\times$0.15 mm$^{3}$.
$M$($T$) was measured using field-cooled (FC) and zero-field-cooled (ZFC) procedures under an external field of 0.4 mT.
Electrical resistivity $\rho$ measurements were performed using a four-probe methodology facilitated by a Quantum Design MPMS apparatus.
In the $\rho$ measurements, we used parallelepiped samples with dimensions of $\sim$5$\times$1$\times$1 mm$^{3}$.

The theoretical underpinning of the electronic structure was achieved through coherent potential approximation (CPA). To this end, the Akai-KKR program package\cite{Akai:JPSJ1982} rooted in the Korringa-Kohn-Rostoker (KKR) method with CPA is harnessed.
We used the generalized gradient approximation from Perdew-Burke-Ernzerhof (PBE) and treated the spin-orbit interaction.

\section{Results and Discussion}\label{sec3}

\begin{figure}
\begin{center}
\includegraphics[width=1\linewidth]{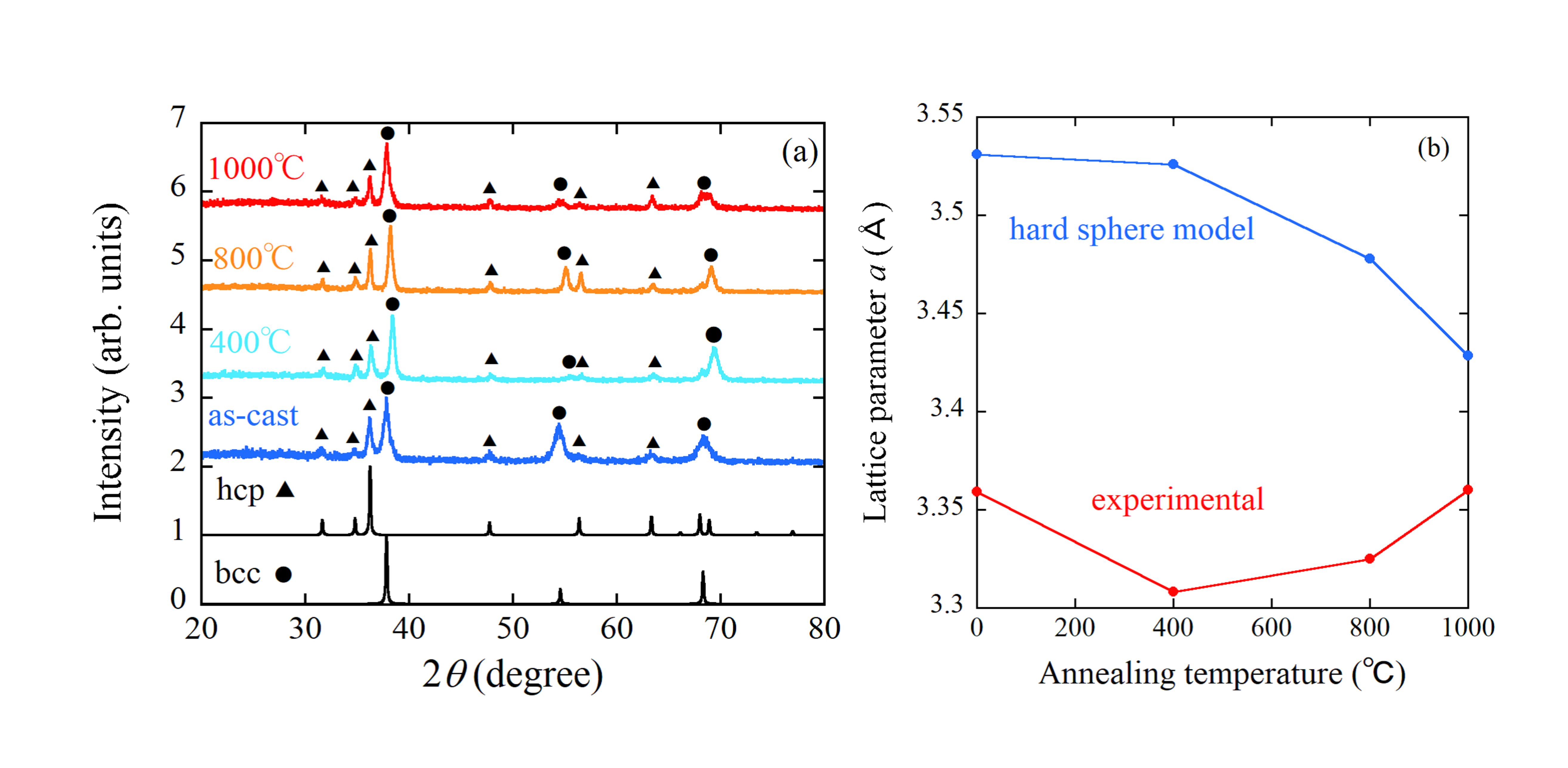}
\caption{\label{fig1}(a) XRD patterns of NbScTiZr annealed at 400 $^{\circ}$C, 800 $^{\circ}$C, and 1000 $^{\circ}$C, together with as-cast sample. The simulation patterns of the bcc ($a$=3.359 \AA) and hcp ($a$=3.262 \AA, $c$=5.152 \AA) structures are also displayed. The origin of each pattern is shifted by a value for clarity. (b) Annealing temperature dependence of the lattice parameters of NbScTiZr (filled red circles). The annealing temperature dependent on the lattice parameters calculated using the hard sphere model is also shown (filled blue circles).}
\end{center}
\end{figure}

Figure \ref{fig1}(a) shows the XRD patterns of the as-cast and heat-treated NbScTiZr samples.
Simulation patterns of the bcc ($a$=3.359 \AA) and hcp ($a$=3.262 \AA, $c$=5.152 \AA) structures were juxtaposed.
All experimental patterns show the concurrent presence of bcc and hcp phases, denoted by filled circles and triangles, respectively.
By employing the least-squares method, the lattice parameters for all the phases were obtained, as listed in Table \ref{tab1}.
In both the bcc and hcp phases, annealing at 400 $^{\circ}$C reduced each lattice parameter.
However, each lattice parameter increased with increasing heat treatment temperature above 400 $^{\circ}$C (see also Fig.\ref{fig1}(b)).

\begin{table}[h]
\caption{Chemical compositions of bcc and hcp phases, volume fractions, lattice parameters, VEC, and $T_\mathrm{c}$ of prepared samples.}\label{tab1}%
\begin{tabular}{@{}llllll@{}}
\toprule
NbScTiZr    & composition   & vol.\%  & lattice parameter (\AA) & VEC & $T_\mathrm{c}$ (K)  \\
\midrule
as-cast    & bcc:Nb$_{26.6(4)}$Sc$_{22.3(5)}$Ti$_{25.5(5)}$Zr$_{25.6(4)}$ & 57  & $a$=3.359(2) & 4.04 & 7.9   \\
           & hcp:Nb$_{21.3(4)}$Sc$_{30(1)}$Ti$_{23.0(7)}$Zr$_{25.8(5)}$  & 43  & $a$=3.262(3), $c$=5.152(3)  & 3.91 &   \\
400 $^{\circ}$C   & bcc:Nb$_{27.9(3)}$Sc$_{21.6(2)}$Ti$_{25.3(3)}$Zr$_{25.2(2)}$  & 57 & $a$=3.308(2) & 4.06 & 8.4  \\
           &hcp:Nb$_{22.1(5)}$Sc$_{32.3(9)}$Ti$_{21.3(5)}$Zr$_{24.2(2)}$  &  43 & $a$=3.252(1), $c$=5.131(1) & 3.90 &   \\
800 $^{\circ}$C    & bcc:Nb$_{33(1)}$Sc$_{15(1)}$Ti$_{32(1)}$Zr$_{20(1)}$  & 50 & $a$=3.325(1)  & 4.18 & 9.0  \\
           &hcp:Nb$_{13(1)}$Sc$_{39.1(9)}$Ti$_{15.1(4)}$Zr$_{32.6(6)}$
       & 50  & $a$=3.252(2), $c$=5.138(3)  & 3.74 &   \\
1000 $^{\circ}$C    & bcc:Nb$_{39.0(8)}$Sc$_{4.2(6)}$Ti$_{38.0(6)}$Zr$_{18.9(8)}$ & 50 & $a$=3.360(1)  & 4.35 & 8.7  \\
           & hcp:Nb$_{2.6(7)}$Sc$_{56(1)}$Ti$_{7.3(7)}$Zr$_{34.1(5)}$ &  50 & $a$=3.260(1), $c$=5.141(2)    & 3.47 &   \\
\botrule
\end{tabular}
\end{table}

\begin{figure}
\begin{center}
\includegraphics[width=1\linewidth]{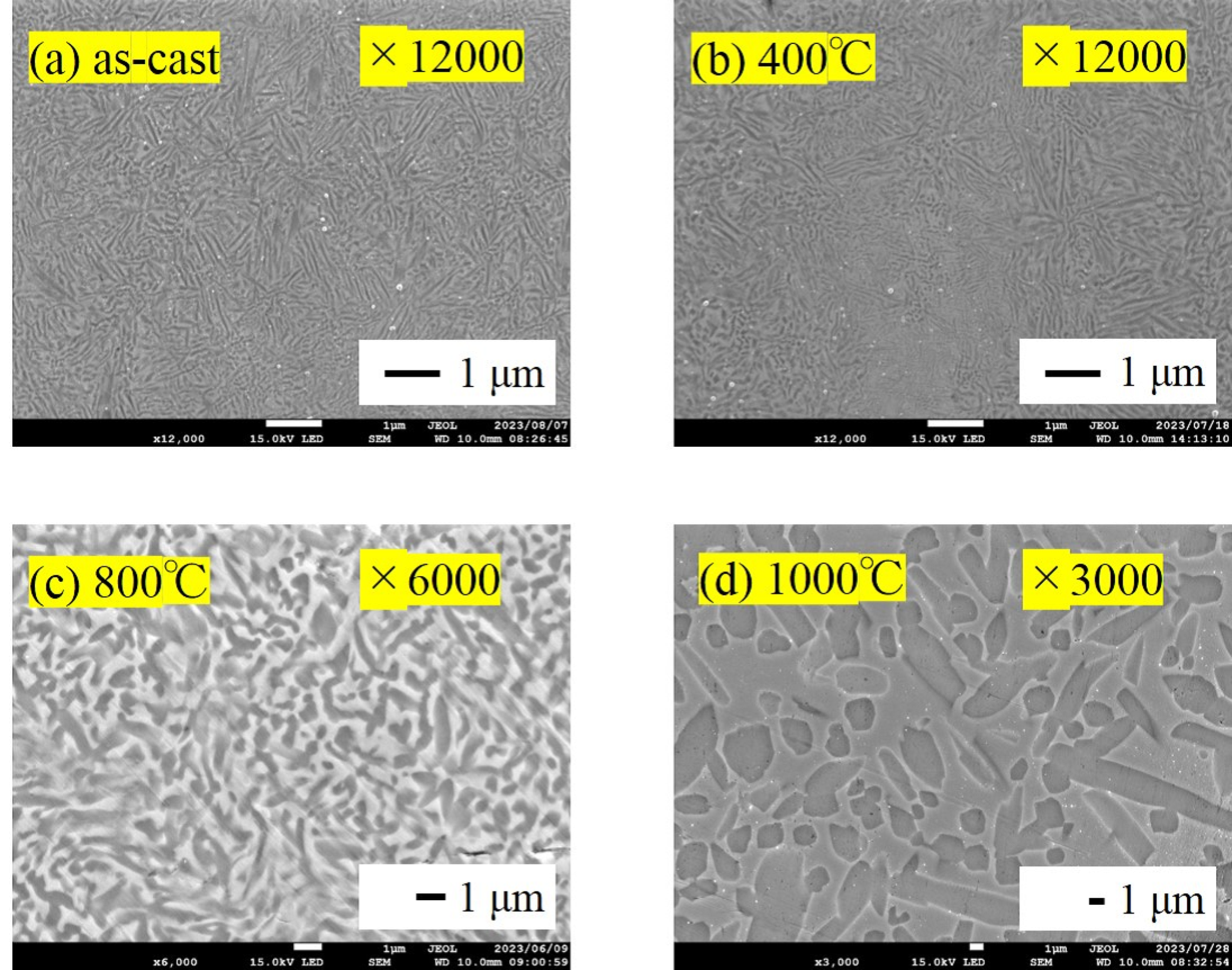}
\caption{\label{fig2} SEM images of NbScTiZr for as-cast sample (a) and heat-treated samples at (b) 400 $^{\circ}$C, (c) 800 $^{\circ}$C, and (d) 1000 $^{\circ}$C, respectively.}
\end{center}
\end{figure}

The SEM images of all samples are shown in Figs. \ref{fig2}(a)-(d).
The as-cast NbScTiZr sample exhibits a lamellar-like arrangement characterized by an approximate thickness of $\sim$ 70 nm ( Fig.\ref{fig2}(a)).
This fine microstructure is consistent with the results of two previous studies \cite{Rogal:MSEA2016,Krnel:Materials2022}.
According to the reports by these research groups, the bright and dark phases were bcc and hcp, respectively.  
The fine structure remained resolute even after 400 $^{\circ}$C annealing (see Fig. \ref{fig2}(b)).
By contrast, when subjected to annealing at higher temperatures, the grain size increased noticeably, as shown in Figs.\ref{fig2}(c) and (d).
With an increase in the annealing temperature beyond 400 $^{\circ}$C, a systematic increase in the grain size is observed, attaining dimensions exceeding 1 $\mu$m after 1000 $^{\circ}$C annealing.
The volume fractions associated with bcc and hcp phases across all samples were obtained from SEM images and are summarized in Table \ref{tab1}.
The ratio of the volume fractions of the bcc and hcp phases was 57:43 in the as-cast sample and its heat-treated counterpart at 400 $^{\circ}$C.
For the samples annealed at 800 $^{\circ}$C and 1000 $^{\circ}$C, the bcc phase fraction exhibited a slight reduction.
The chemical compositions determined by EDX for the bcc and hcp phases in each sample are summarized in Table \ref{tab1}. 
In the as-cast sample, the bcc and hcp phases emerged as Sc-deficient and Sc-rich, respectively.
This chemical composition maintained a nearly identical state, even within the sample annealed at 400 $^{\circ}$C.
Profound shifts in the chemical compositions of the bcc and hcp phases occur within the samples annealed at 800 $^{\circ}$C and 1000 $^{\circ}$C; the bcc phase shows increased Nb and Ti contents, whereas the hcp phase enriches the Sc and Zr contents.
Drawing on the insights gained from the chemical analysis, we discuss the annealing temperature dependence of the lattice parameters of the bcc phase. 
We invoked the framework of the hard-sphere model in conjunction with the obtained chemical composition, enabling the estimation of the ideal lattice parameter across distinct annealing conditions.
By employing atomic radii of 1.429 \AA \hspace{1mm} for Nb, 1.641 \AA \hspace{1mm} for Sc, 1.4615 \AA \hspace{1mm} for Ti, and 1.6025 \AA \hspace{1mm} for Zr\cite{Miracle:AM2017}, application of the hard sphere model yields the lattice parameter, which is expressed as 2.309$\bar{r}$, where $\bar{r}$ is the composition-weighted atomic radius. 
The resultant lattice parameters were compared with experimental data, as shown in Fig. \ref{fig1}(b), wherein the as-cast sample is regarded as the analog for 0 $^{\circ}$C annealing.
The calculated lattice parameter closely approximates its experimental counterpart at 1000 $^{\circ}$C, implying a weak influence of lattice strain, probably owing to the annealing effect.
By contrast, at lower temperatures, such as the as-cast and 400 $^{\circ}$C annealing conditions, a substantial disparity between the calculated and experimental lattice parameters becomes evident, suggesting a plausible manifestation of lattice strain.

\begin{figure}
\begin{center}
\includegraphics[width=1.1\linewidth]{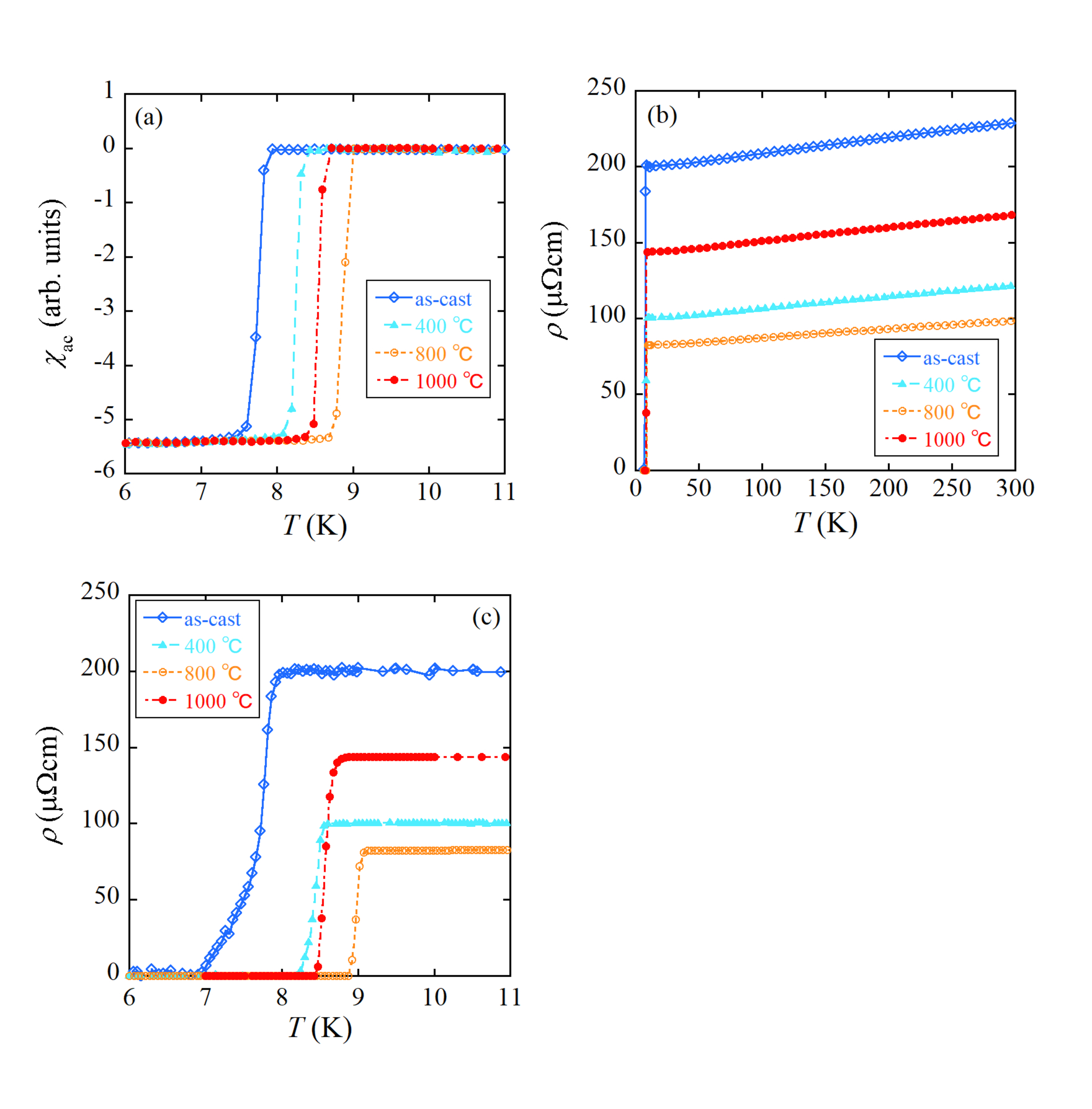}
\caption{\label{fig3} Temperature dependences of (a) $\chi_{ac}$, (b) $\rho$, and (c) low temperature $\rho$ of NbScTiZr.}
\end{center}
\end{figure}

Figure \ref{fig3}(a) shows $\chi_{ac}$ ($T$) values of all the samples. 
The onset of the diamagnetic signal is denoted by $T_\mathrm{c}$.
The as-cast sample showed a diamagnetic signal below $T_\mathrm{c}$=7.9 K, which agrees with the findings of Krnel et al.\cite{Krnel:Materials2022}.
Remarkably, our investigations revealed a progressive increase in $T_\mathrm{c}$ to 9.0 K with increasing annealing temperatures up to 800 $^{\circ}$C.
However, a subsequent increase in the annealing temperature yielded a marginal decline in $T_\mathrm{c}$ to 8.7 K.
We also confirmed a zero $\rho$ value for each sample, as shown in Figs. \ref{fig3}(b) and (c).
Importantly, although the order of magnitude characterizing the normal state $\rho$ for each sample suggests a metallic nature, atomic disorder contributes to the weak temperature dependence.
In each sample, $\rho$ drops below approximately $T_\mathrm{c}$, as determined by $\chi_{ac}$ ($T$) measurement. 
Although the superconductive transition in $\rho$ ($T$) of the as-cast sample was relatively broad, it became sharper after annealing ( Fig. \ref{fig3}(c)).

\begin{figure}
\begin{center}
\includegraphics[width=0.6\linewidth]{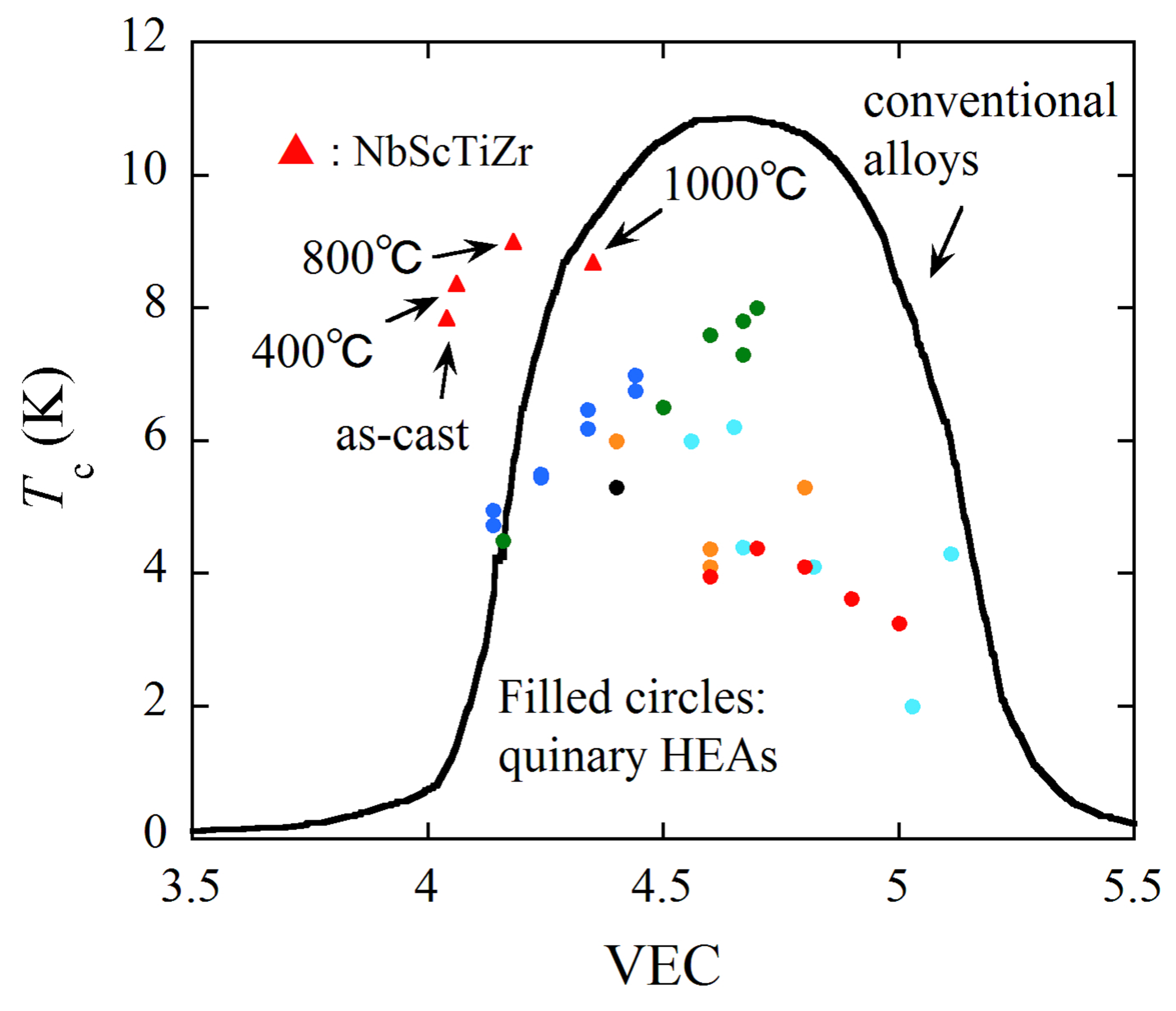}
\caption{\label{fig4} VEC dependence of $T_\mathrm{c}$ of NbScTiZr (red triangle). The solid line represents the Matthias rule of conventional binary or ternary transition metal alloys. The data of typical quinary bcc HEA superconductors are also shown by the filled circles. For the quinary bcc HEAs, the correspondence between color and HEA is as follows; green: nonequimolar Hf-Nb-Ta-Ti-Zr, blue: Al-Nb-Ti-V-Zr, black: Hf$_{21}$Nb$_{25}$Ti$_{15}$V$_{15}$Zr$_{24}$, orange: HfNbTaTiZr, HfNbReTiZr, HfNbTaTiV, and HfMoNbTiZr, light blue: Nb-Ta-Mo-Hf-W, Ti-Zr-Nb-Ta-W, and Ti-Zr-Nb-Ta-V, and red: Ti-Hf-Nb-Ta-Re.}
\end{center}
\end{figure}

\begin{figure}
\begin{center}
\includegraphics[width=1.0\linewidth]{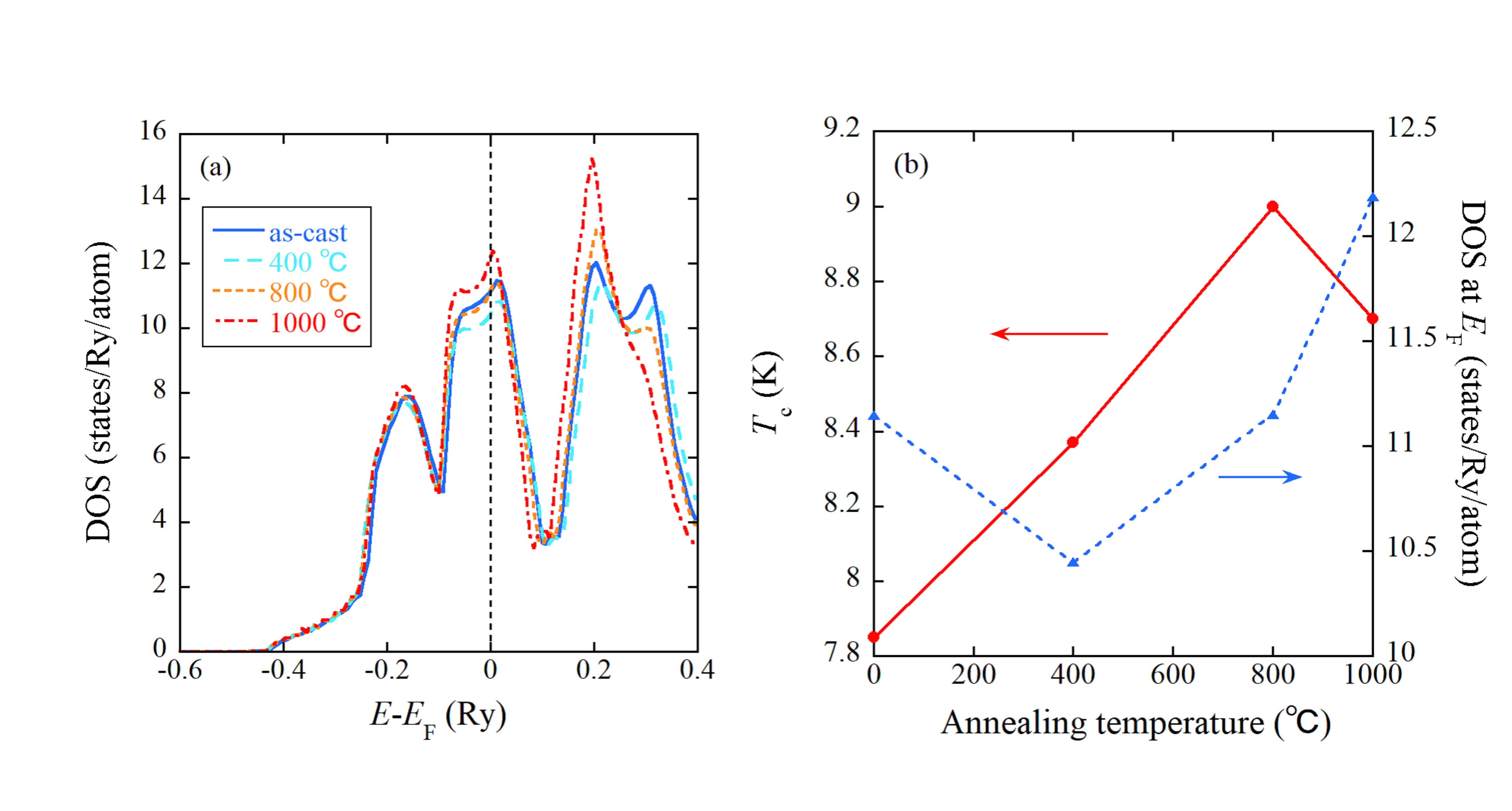}
\caption{\label{fig5} (a) Total electronic density of states for NbScTiZr. (b) Annealing temperature dependence of $T_\mathrm{c}$ and $D(E_\mathrm{F})$.}
\end{center}
\end{figure}

Herein, we discuss the interplay between valence electron concentration (VEC) and $T_\mathrm{c}$.
Within the framework of the BCS theory, $T_\mathrm{c}$ depends on the density of states (DOS) at the Fermi level ($E_\mathrm{F}$), denoted as $D(E_\mathrm{F})$, and the electron-phonon interaction. 
The VEC reflects $D(E_\mathrm{F})$, and a strong correlation between the VEC and $T_\mathrm{c}$ has been observed in binary or ternary superconducting transition metal alloys, referred to as the Matthias rule\cite{Matthias:PR1955}. 
The computational derivation of VEC adopts the following equation: 
\begin{equation}
\mathrm{VEC}=\sum^{4}_{i=1}c_{i}\mathrm{VEC}_{i}
\label{vec}
\end{equation}
where $c_{i}$ and VEC$_{i}$ are the molar fraction and VEC value of the $i$th element, respectively. 
The specific VEC$_{i}$ values are 5 for Nb, 3 for Sc, and 4 for Ti and Zr.
Considering the chemical composition of the bcc phase in each sample, the relationship between $T_\mathrm{c}$ and VEC for NbScTiZr samples is shown in Fig.\ref{fig4} (see also Table \ref{tab1}).
The solid line in the figure describes the Matthias rule intrinsic to conventional binary or ternary transition metal alloys\cite{Matthias:PR1955}, revealing a broad peak at a VEC value of approximately 4.6.
The dataset of typical quinary bcc HEA superconductors is also shown by filled circles (green: nonequimolar Hf-Nb-Ta-Ti-Zr, blue: Al-Nb-Ti-V-Zr, black: Hf$_{21}$Nb$_{25}$Ti$_{15}$V$_{15}$Zr$_{24}$, orange: HfNbTaTiZr, HfNbReTiZr, HfNbTaTiV, and HfMoNbTiZr, light blue: Nb-Ta-Mo-Hf-W, Ti-Zr-Nb-Ta-W, and Ti-Zr-Nb-Ta-V, and red: Ti-Hf-Nb-Ta-Re.)\cite{Marik:JALCOM2018,Ishizu:RINP2019,Harayama:JSNM2021,Sarkar:IM2022,Kitagawa:JALCOM2022,Hattori:JAMS2023,Vrtnik:JALCOM2017,Rohr:PNAS2016,Sobota:PRB2022,Shu:APL2022}. 
Conventionally, the dataset of HEAs traces a trajectory beneath a solid line.
NbScTiZr annealed at 1000 $^{\circ}$C resides near the solid line.
The sample annealed at 800 $^{\circ}$C, despite having a diminished VEC value compared with its 1000 $^{\circ}$C counterpart, experiences an $T_\mathrm{c}$ enhancement that is inconsistent with the anticipation of the Matthias rule.
A further strange observation emerges for the as-cast NbScTiZr and the sample subjected to 400 $^{\circ}$C annealing, which deviates from the solid line and HEAs trend. 
The Matthias rule is based on an elevated $D(E_\mathrm{F})$ at a VEC of approximately 4.6, accompanied by a relatively low $D(E_\mathrm{F})$ on either side of the threshold. 
Consequently, at a VEC of approximately 4.0, a considerable decline in $D(E_\mathrm{F})$ was predicted, causing a notable decrease in $T_\mathrm{c}$.
To explore the underpinnings of $D(E_\mathrm{F})$ in NbScTiZr, electronic structure calculations were performed, the results of which are shown in Figs.\ref{fig5}(a), offering DOS for NbScTiZr across all annealing conditions.
The experimental lattice parameters and chemical compositions determined by EDX were used for the electronic structure calculations.
In as-cast NbScTiZr or the sample annealed at 400 $^{\circ}$C, $D(E_\mathrm{F})$ stands for analogous bcc HEAs\cite{Sarkar:IM2022,Kitagawa:JALCOM2022,Jasiewicz:PSSRRL2016}, suggesting the plausible emergence of superconductivity with $T_\mathrm{c}$ higher than that predicted by the Matthias rule.
We can extract the value of $D(E_\mathrm{F})$ for each annealing condition and plot it as a function of the annealing temperature in Fig.\ref{fig5}(b).
As the annealing temperature increased beyond 400 $^{\circ}$C, $D(E_\mathrm{F})$ exhibited a systematic increase, driven by the concurrent increase in VEC. 
Moreover, when transitioning from the as-cast sample to its 400 $^{\circ}$ C-annealed counterpart, a reduction in $D(E_\mathrm{F})$ was observed.
Fig.\ref{fig5}(b) further elucidates the annealing temperature-sensitive $T_\mathrm{c}$. 
Although we expect an enhancement in $T_\mathrm{c}$ as $D(E_\mathrm{F})$ increases, such a relationship is partially observed.
It is evident that factors beyond $D(E_\mathrm{F})$ interact to delineate the annealing temperature dependence of $T_\mathrm{c}$, which suggests the potential role of the electron-phonon interaction.
As evidenced by Fig.\ref{fig1}(b), the lattice parameter shown in the 1000 $^{\circ}$C annealing agrees with that predicted by the hard sphere model. This result underscores the weak influence of lattice strain. 
Conversely, for the other samples, conspicuous disparities between the experimental and hard-sphere model-derived lattice parameters manifested, signaling the presence of lattice strain. 
Such an occurrence could conceivably contribute to alterations in the electron-phonon interactions, subsequently influencing the behavior of $T_\mathrm{c}$.
Notably, the lattice strain occurring at annealing temperatures below 800 $^{\circ}$C may be partially responsible for the enhancement in $T_\mathrm{c}$, surpassing the value predicted by the Matthias rule.
This phenomenon resembles the enhancement of $T_\mathrm{c}$ reported for Sr$_{2}$RuO$_{4}$, Ir, and Zr$_{5}$Pt$_{3}$O$_{x}$, as explained in Introduction.

We discuss the impact of annealing on bcc HEA superconductors by drawing upon a comparative analysis with relevant literature.
Two notable studies extensively investigated Hf-Nb-Ta-Ti-Zr HEA superconductors\cite{Vrtnik:JALCOM2017,Gao:APL2022}. 
In one of these studies, three distinct HEAs, each characterized by unique chemical composition (Hf$_{20}$Nb$_{21}$Ta$_{20}$Ti$_{19}$Zr$_{20}$, Hf$_{21}$Nb$_{24}$Ta$_{22}$Ti$_{10}$Zr$_{23}$, and Hf$_{26}$Nb$_{25}$Ta$_{25}$Zr$_{24}$), are subjected to annealing procedures within the temperature range of 1800 $\sim$ 2000 $^{\circ}$C\cite{Vrtnik:JALCOM2017}. 
Remarkably, the superconducting transition manifests robustly during thermal annealing, although the $T_\mathrm{c}$ value depends on chemical composition.
Notably, although the HEAs exhibit no eutectic structures, each sample exhibits a dendritic microstructure. 
Nevertheless, the HEAs did not show a definitive correlation between microstructural morphology and $T_\mathrm{c}$.
In the other study\cite{Gao:APL2022}, the annealing effect of $T_\mathrm{c}$ in (TaNb)$_{0.7}$(HfZrTi)$_{0.5}$ was explored.
At annealing temperatures exceeding 700 $^{\circ}$C, the phase segregation becomes apparent. 
$T_\mathrm{c}$ of approximately 7 K remained nearly unchanged upon annealing, even after the occurrence of phase segregation.
Furthermore, precipitation phenomena emerged above an annealing threshold of 500 $^{\circ}$C, serving as effective flux pinning sites, as subsequently delineated. 
However, $T_\mathrm{c}$ was insensitive to alterations in the microstructure. 
In NbScTiZr, the change in chemical composition depending on the annealing temperature partially affected the $T_\mathrm{c}$ value, paralleling the findings of a previous report\cite{Vrtnik:JALCOM2017}.
Notably, the influence of the eutectic structures on $T_\mathrm{c}$, mediated by the lattice strain, was also confirmed in the NbScTiZr system. 
This starkly contrasts with Hf-Nb-Ta-Ti-Zr HEA superconductors, which exhibit a conspicuous insensitivity to alterations in microstructure in the context of $T_\mathrm{c}$. 
Hf-Nb-Ta-Ti-Zr HEA superconductors are devoid of lattice strain. 
Consequently, it becomes evident that the lattice strain plays a pivotal role in the systematic variation of $T_\mathrm{c}$ during the thermal annealing.

\begin{figure}
\begin{center}
\includegraphics[width=1\linewidth]{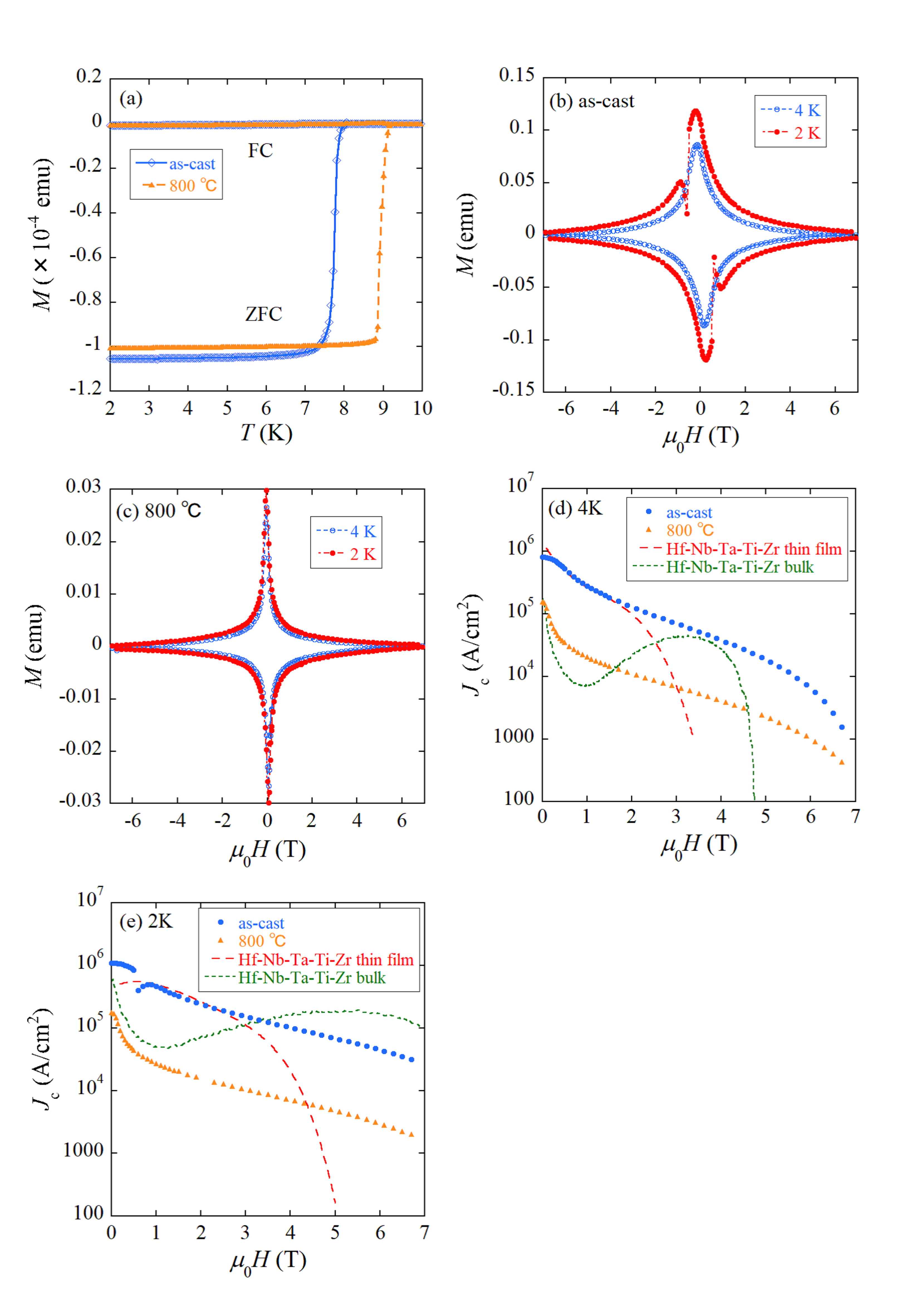}
\caption{\label{fig6} (a) Temperature dependence of magnetization of NbScTiZr. Field-dependent magnetization of NbScTiZr at 2 and 4 K for (b) as-cast sample and (c) sample annealed at 800 $^{\circ}$C. Field-dependent $J_\mathrm{c}$ of as-cast NbScTiZr and sample annealed at 800 $^{\circ}$C at (d) 4 K and at (e) 2 K. In (d) and (e), the data\cite{Jung:NC2022,Gao:APL2022} of representative bcc HEA superconductors are also shown.}
\end{center}
\end{figure}

Figure \ref{fig6}(a) shows $M$($T$) under an external field of 0.4 mT for both the as-cast NbScTiZr and sample annealed at 800 $^{\circ}$C.
In each sample, $M$($T$) measured under ZFC conditions exhibited a diamagnetic signal, whereas the $M$($T$) data obtained under FC conditions indicated the manifestation of flux pinning within the sample.
The onset temperatures of the diamagnetic signals were consistent with $T_\mathrm{c}$ values determined from the $\chi_{ac}$ ($T$) measurements. 
The results of the isothermal hysteresis loops in $M$–$H$ ($H$: external field) curves at both 2 and 4 K are shown in Figs.\ref{fig6}(b) and (c) for the as-cast NbScTiZr and the sample annealed at 800 $^{\circ}$C, respectively.
Both samples exhibited hysteresis loops that are characteristic of type-II superconductors. 
The hysteresis loop corresponding to the sample annealed at 800 $^{\circ}$C demonstrates a small area relative to the as-cast counterpart.
For the as-cast sample at 2 K, abrupt changes in $M$ occur at approximately 5 kOe during the field-increasing process and at -5 kOe during the field-decreasing process, as shown in Fig.\ref{fig6}(b).
By contrast, no such abrupt variations in $M$ were discernible at 4 K, as shown in the same figure.
This behavior is often observed in several superconductors, including HEAs\cite{Jung:NC2022,Gao:APL2022,Dou:PhysicaC2001}.
Abrupt modulation of $M$ is commonly attributed to the flux jump phenomenon, which is indicative of pronounced local magnetic instability.

$J_\mathrm{c}$ is assessed by using the critical state model expressed as follows:\cite{Peterson:JAP1990}
\begin{equation}
J_\mathrm{c}=\frac{3\Delta M}{d}
\label{Jc}
\end{equation}
where $\Delta M$ is the width of the loop at applied $H$ and $d$ represents the effective sample width perpendicular to $H$.
Thus, the calculated $H$ dependence of $J_\mathrm{c}$ is shown in Figs.\ref{fig6}(d) and (e).
$J_\mathrm{c}$ exhibited an ascending trend with decreasing temperature in both the as-cast and heat-treated samples.
Notably, the self-field $J_\mathrm{c}$ of the as-cast sample at 2 K exceeded 10$^{6}$ A/cm$^{2}$, classifying it as a material with a high $J_\mathrm{c}$. 
Intriguingly, the $J_\mathrm{c}$ value of the as-cast sample was higher than that of the sample annealed at 800 $^{\circ}$C by an order of magnitude under fixed $H$ conditions. 
This enhanced $J_\mathrm{c}$ signifies pronounced magnetic-flux pinning.
As observed in Figs.\ref{fig2}, the eutectic microstructure of the as-cast sample is distinctly fine, in contrast to the coarser morphology evident in the sample annealed at 800 $^{\circ}$C. 
Typically, the grain boundaries serve as loci for flux pinning. 
The fine eutectic structure in the as-cast sample is endowed with a greater abundance of grain boundaries between the bcc and hcp phases compared to the sample annealed at 800 $^{\circ}$C, which is characterized by a coarser microstructure. 
This difference accounts for the elevated $J_\mathrm{c}$ of the as-cast sample, indicating the pivotal role of the fine eutectic structure in enhancing $J_\mathrm{c}$.
Figures \ref{fig6}(d) and (e) also show data from a Hf-Nb-Ta-Ti-Zr bcc HEA film, which exhibits the highest self-field $J_\mathrm{c}$ among bcc HEA superconductors\cite{Jung:NC2022}. 
Furthermore, $J_\mathrm{c}$ data from a bulk Hf-Nb-Ta-Ti-Zr ((TaNb)$_{0.7}$(HfZrTi)$_{0.5}$) bcc HEA sample were included\cite{Gao:APL2022}. 
Although the self-field $J_\mathrm{c}$ of this bulk HEA is lower by order of magnitude than that of the Hf-Nb-Ta-Ti-Zr film, it demonstrates relatively higher $J_\mathrm{c}$ values under higher fields ($\mu_{0}H$ $>$ 3 T), which is attributable to the fishtail effect\cite{Gao:APL2022}. 
The fishtail effect is ascribed to precipitation during the heat treatment process, which enhances flux pinning.
Notably, at both 2 and 4 K, $J_\mathrm{c}$ of the as-cast NbScTiZr sample closely aligns with that of the Hf-Nb-Ta-Ti-Zr bcc HEA film below 2 T and surpasses $J_\mathrm{c}$ of the film above 2-3 T. 
At elevated applied fields, the $J_\mathrm{c}$ of the as-cast NbScTiZr becomes competitive with that of the Hf-Nb-Ta-Ti-Zr bulk bcc HEA. 
These comparative assessments highlight the potential of eutectic HEA superconductors to exhibit high $J_\mathrm{c}$ values.
Herein, we briefly discuss a comparative analysis of the field-dependent $J_\mathrm{c}$ behaviors of NbScTiZr and (TaNb)$_{0.7}$(HfZrTi)$_{0.5}$.
In the (TaNb)$_{0.7}$(HfZrTi)$_{0.5}$ system, the precipitate takes the form of clusters characterized by dimensions of the order of $\sim$ 10 nm\cite{Gao:APL2022}. 
This feature confers a pronounced advantage, particularly in the augmentation of $J_\mathrm{c}$ performance at elevated magnetic fields. 
Conversely, in NbScTiZr, we encountered a fine eutectic microstructure that substantially supported the enhancement of $J_\mathrm{c}$ at lower magnetic fields.
Hence, the synergy achieved through the presence of both the eutectic microstructure and cluster-like nanosized precipitates may realize elevated $J_\mathrm{c}$ values over a wide field range.

\section{Summary}\label{sec4}
We found that the $T_\mathrm{c}$ of the eutectic HEA superconductor NbScTiZr systematically increases from 7.9 to 9 K with increasing thermal annealing temperatures up to 800 $^{\circ}$C.
Annealing at 1000 $^{\circ}$C slightly reduces $T_\mathrm{c}$. 
Analysis of the lattice parameters using the hard-sphere model indicates the presence of potential lattice strain, particularly at lower annealing temperatures.
While the lamellar-like structure with a thickness of approximately 70 nm endured annealing up to 400 $^{\circ}$C , a systematic enlargement of the grain size occurred with an increase in the annealing temperature beyond 400 $^{\circ}$C.
We discussed the annealing temperature dependence of $T_\mathrm{c}$ by employing the $T_\mathrm{c}$ vs. VEC plot and electronic structure calculations.
Factors other than $D(E_\mathrm{F})$ should be considered to explain the annealing-temperature dependence of $T_\mathrm{c}$.
Notably, the electron-phonon interactions are pivotal in this regard. 
Furthermore, the conceivable influence of the lattice strain, deliberated within the context of the annealing-temperature dependence of the lattice parameter, may affect the annealing-temperature dependence of $T_\mathrm{c}$. 
For further exploration, we assessed $J_\mathrm{c}$ for both the as-cast NbScTiZr and sample annealed at 800 $^{\circ}$C and classified the as-cast NbScTiZr as a material with a notably high $J_\mathrm{c}$.

\bmhead{Acknowledgments}

T.N. acknowledges the support from a Grant-in-Aid for Scientific Research (KAKENHI) (Grant No. 20K03867) and the Takahashi Industrial and Economic Research Foundation. Y.M. acknowledges the support from a Grant-in-Aid for Scientific Research (KAKENHI) (Grant No. 21H00151). J.K. acknowledges the support from a Grant-in-Aid for Scientific Research (KAKENHI) (Grant No. 23K04570) and the Comprehensive Research Organization of the Fukuoka Institute of Technology.

\end{document}